# Heuristics in experiments with infinitely large strategy spaces

Jørgen Vitting Andersen♣,♣♣, and Philippe de Peretti♣♣

♣ CNRS, ♣♣Centre d'Economie de la Sorbonne, Université Paris 1 Pantheon-Sorbonne, Maison des Sciences Economiques,106-112 Boulevard de l'Hôpital, 75647 Paris Cedex 13, France.

**We introduce a new methodology that enables detection of the onset of convergence towards Nash equilibria in simple repeated games with infinitely large strategy spaces, thereby revealing the heuristics used in decision-making. The method works by constraining on a special finite subset of strategies, called decoupled strategies. We show how the technique can be applied to understand price formation in financial market experiments by introducing a predictive measure $\Delta D$: the different between positive decoupled strategies (recommending to buy) and negative decoupled strategies (recommending to sell). Using $\Delta D$ we illustrate how the method can predict (at certain special times) participants' actions with a high success rate in a series of experiments.**

## 1 Introduction

It is often difficult to obtain a proper behavioral prediction of people's actions in simple one-shot games, since they are obscured by non-equilibrium behavior which can persist even when the games are played repeatedly (*Chong et al. 2016; Muller and Tan 2013*). One question then is: how can we get a better understanding of the decision-making in repeated games, and if possible the dynamics thereof, before a proper equilibrium has set in? In *Chong et al. (2016)*, a generalized cognitive hierarchy was introduced to capture the fact that subjects frequently do not play a Nash equilibrium in simple one-shot games, a behavioral trait which could arise because players are heterogeneous and have different thinking abilities. As will be explained below, the fact that there are different levels of thinking ability, and the impact this could have on convergence to equilibrium, is an important issue in both the experiments and the modeling described in this article. At the same time, it illustrates a method to understand the heuristics people use in experiments where the decision-making becomes increasingly complicated.

The situation we describe here is complex since we are considering the case when the available strategy space becomes infinitely large. In situations like this it is unrealistic to think that people analyze an infinite range of strategies. Rather, it becomes more likely that people use heuristics *(Crawford 2013)*, limiting the analyses to a few simple and understandable cases. Then the question is how can we try to model such an analysis in a game theoretical framework?

Here we suggest one solution, which is to focus on a certain subset of strategies. The solution is reminiscent of the study by *Ioannou and Romeo (2014)*, who introduced a methodology to facilitate the operability of belief-learning models with repeated-game strategies. Since the set of possible strategies in repeated games is infinite (uncountable), expecting a player to fully explore such an infinite set is unrealistic. As noted in *Ioannu and Romeo (2014)*, and in *McKelvey and Palfrey (2001)*, there is also an inference problem, even for games with only two players. In repeated games there is no unique way to identify an opponent's strategy just on the basis of the history, since several different strategies can lead to the same history. This point becomes even more relevant for the type of games we present in this study, where the history is created by several (N>2) players, making it impossible for a given player to know exactly what strategies other players have adopted. Therefore, even though the history of play is publicly observable, a player will never be able to deduce an opponent's strategy in any detail.

The solution proposed by *Ioannu and Romeo (2014)* was to introduce repeated-game strategies implemented via a type of finite automaton, called a Moore machine, thereby limiting the infinite repeated-game strategy space to a finite subset. In our study we will focus on another special subset of repeated-game strategies, called decoupled strategies, which have the special property that their actions are *independent* of the action history, *conditioned* on observing certain histories. As will be seen, this property allows us to identify heuristic strategies automatically, since at special moments the decoupled repeated–game strategies then effectively correspond to simple one-shot strategies. We would also like to point out the similarity with collective action games; see for example *Anauati et al. (2016)*, who introduced a method using stability sets. They were able to capture the case where increases in the payoff of a successful collective action led to an update in prior beliefs about the expected share of cooperators. As will be seen, we encounter similar situations where increases in the payoff of a successful aggregate trading action, leads to an increase in the probability of the use of certain trading strategies.

As will be explained below, in the following we will use the repeated minority game (or more precisely: a slightly modified version called the $-game) as a case study for this paper. The definition of the minority game is presented in the section below, it was introduced by *Challet and Zhang (1997)* as a simple model to describe price formation in financial markets. Some of the first analytical results on the model was made by *Challet et al (2000b)*, see also *Challet and Zhang (1998)*. *Lamper et al. (2001)* were the first to point out that the minority game had the particular property of showing predictability of large future price changes. The mechanism leading to such large price changes were however not identified in *Lamper et al. (2001)*. *Andersen and Sornette (2005)* introduced a mechanism that could describe such "pockets of predictability", by considering a certain class of strategies, called decoupled strategies. In the following we will use such decoupled strategies to identify the heuristics used by the participants in experiments on financial markets. For some of the few experimental efforts that has been done to understand the minority game, a nice and general multi-round strategy experiment on the minority game can be found in *Linde et al. (2014)*.

## 2 Simple multi-period market games: theoretical framework

When the economist Brian Arthur introduced his famous *El Farol* bar game (*Arthur 1994)*, it set off a flurry of research on simple binary choice decision-making models, with decisions like "go/stay", "yes/no", "buy/sell", "right/left", etc. In Arthur's model, a population of N=100 people want to go to the El Farol bar (a bar that really exists in Sante Fe) to listen to folk music. However, the El Farol is quite small, and people are only satisfied if they can actually sit on one of the 60 available seats. People have to make their decision at the same time, so if everybody uses the same pure strategy, it will fail: suggesting "go" (assuming an empty bar), everybody will go and the bar will be crowded, but suggesting "stay" (assuming a crowded bar) will instead lead to an empty bar. *Challet and Zhang (1997)* extended the El Farol model to describe financial market behavior, where the binary choice was now a simple "buy/sell" decision. More precisely, in the minority game, there is an odd number N of players who use different strategies to try to remain on the minority side. If a strategy $S_i$ predicts that the majority will buy, then that strategy will recommend to sell, $S_i = -1$. Otherwise, if it predicts that the majority will sell, the recommendation will be to buy, $S_i = 1$. The payoff of strategy *i* is

$$\pi^{MG}(s_i) = -s_i A, \quad A = \sum_{j}^{N} s_j$$

where *A* is the order imbalance (i.e., the difference in buy and sell orders). As mentioned in *Linde et al. (2014)*, the one-shot minority game already has quite a large number of Nash equilibria. This happens since any case where exactly (N-1)/2 players choose one side (and (N+1)/2 players the other side) constitutes a Nash equilibrium. The number $\frac{N!}{\left(\frac{N+1}{2}\right)! \left(\frac{N-1}{2}\right)!}$ of Nash equilibria is already quite large even for moderate values of N.

In the multi-round minority game, players then use strategies that consider the outcome of not only the last, but the past M market price directions (for a positive order imbalance A, the market goes up; when A is negative, the market goes down). Each player holds the same number of strategies S, assigned randomly at the beginning of the game. A strategy in the multi-round game issues a prediction of the next market move (buy/sell) for each of the possible $2^M$ past price histories. A strategy in the multi-round minority game is therefore a vector of size $2^M$ instructing whether to buy (1), or sell (-1) for each possible price history. An example of a multi-round strategy could for example be (1,0,1) → 1, (M=3), which means that in a market where the last price movement was up, the one before down, and the price movement three time steps ago up, it would recommend to buy. The total number of different strategies is therefore $2^{2^M}$. In the multi-round minority game, the players record the cumulative payoff for each of their S strategies, and at any given time, use the one which has the highest cumulative payoff. For an extensive review of the multi-round minority game, see for example (*Cavagna et al. 1999; Challet et al. 2000a; Challet, et al. 2000b; Challet et al. 2001, Challet and Zhang, Y.C., 1998, Johnson et al. 1999; Lamper et al. 2001*).

As mentioned in *Andersen and Sornette (2003)*, one problem with the multi-round minority game describing simple market price dynamics is that it does not account for speculative behavior, in which investors invest to gain a return. It should be noted that the minority dynamics prevents any trend to develop, giving rise to mean-reverting price dynamics. In order to capture speculative behavior as seen in bubble/crash phases of financial markets, a modification of the payoff function was suggested in *Andersen and Sornette (2003)*. Denoting the game, the $-game (to describe players that speculate), the modified payoff reads:

$$\pi^{\$G}(s_i(t)) = s_i(t-1) R(t)$$

The payoff favors strategies which are able to predict the price movement over the following time step. Predicting at time t-1 a price increment over the following time step, the strategy $i$ proposes to enter a buy position at time t-1, so $s_i(t-1) = 1$. If the prediction was successful (a failure), the payoff gained (lost) is the return of the market over that time step.

**Theorem 2.1** *The Nash equilibria in the multi-period $-game consist of the two strategies (1,1,....,1), (-1,-1,...,-1).*

**Proof** We will prove theorem 2.1 for the simple case where each agent has only s=2 strategies. The dynamics of the $-game is driven by a nonlinear feedback mechanism, because each agent uses his/her best strategy (fundamental/technical analysis) at each time step. In its turn, the sign of the order imbalance, $\sum_{i=1}^{N} a_i^*(\vec{h}(t))$, determines the value of the last bit $b(t)$ at time $t$ for the price movement history $\vec{h}(t+1) = (b(t-m+1), b(t-m), ..., b(t))$. The * in $a_i^*$ denotes the best strategy (out of s possible) for agent i. The dynamics of the $-game can then be expressed in terms of an equation that describes the dynamics of $b(t)$ as

$$b(t+1) = \Theta[\sum_{i=1}^{N} a_i^*(h(t))] \tag{1}$$

where $\Theta$ is a Heaviside function taking the value 1 whenever its argument is greater than 0 and the value 0 otherwise, and $h(t) = \sum_{j=1}^{m} b(t-j+1)2^{j-1}$ is now expressed as a scalar instead of a vector. The nonlinearity of the game can be seen formally from

$$a_i^*(h(t)) = a_i^{\{j|max_{j=1,...,s}[\sum_{k=1}^{t} a_i^j(h(k-1))\sum_{i=1}^{N} a_i^*(h(k))]\}}(h(t)) \tag{2}$$

Equation (2) expresses the idea that the optimal strategy * of agent i is the strategy j which maximizes the $G payoff between times k=1 and k=t. Inserting (2) in (1), we obtain an expression that describes the $-game in terms of just a single equation for $b(t)$ depending on the values of the 5 base parameters (m; s; N; λ; D(t)) and the random variables $a_i^j$ (i.e., their initial random assignments).

We would like first to point out an important difference compared to traditional game theory, since in our game the agents have no direct information concerning the action of the other players. The only (indirect) information a given agent can have of another agent's action comes through the aggregate actions of the past, i.e., the past price behavior.

In the following we will consider the simple case where each agent has only s=2 strategies. Then considering only the relative payoff between strategy $a_i^1$ and $a_i^2$, the equations simplify considerably. Let the action of the optimal strategy $a_i^*$ be expressed in terms of the relative payoff, $q_i$, so as to formulate $\sum_{i=1}^{N} a_i^*(h(t))$ as follows:

$$\sum_{i=1}^{N} a_i^*(h(t)) = \sum_{i=1}^{N} \left\{ \Theta\left(q_i(h(t))\right) a_i^2(h(t)) + \left[1 - \Theta\left(q_i(h(t))\right)\right] a_i^1(h(t)) \right\} \quad (3)$$

Inserting (3) into (1) and taking the derivative of $b$ at $t+1$,

$$\left.\frac{db}{dt}\right|_{t+1} = \delta\left(\sum_{i=1}^{N} a_i^*(h(t))\right) \sum_{i=1}^{N} \left\{ \delta\left(q_i(h(t))\right) \frac{\delta q_i(h(t))}{\delta t} \left[a_i^2(h(t)) - a_i^1(h(t))\right] \right.$$

$$\left. + \Theta\left(q_i(h(t))\right) \frac{\delta a_i^2(h(t))}{\delta t} + \left[1 - \Theta\left(q_i(h(t))\right)\right] \frac{\delta a_i^1(h(t))}{\delta t} \right\} \quad (4)$$

Looking at the terms inside the curly bracket in (4), it follows that a change in $\sum_{i=1}^{N} a_i^*(h(t))$ can occur in two different ways: (i) because the optimal strategy changes and the two strategies for a given $h(t)$, $a_i^1(h(t))$ and $a_i^2(h(t))$, differ from each other (first term in the bracket); (ii) because the optimal strategy changes its prediction for the given $h(t)$ (second and third terms in the bracket).

The rate of change of the relative payoff $q_i$ is computed from

$$\left.\frac{\delta q_i}{\delta t}\right|_t = \left[a_i^2(h(t-2)) \sum_{i=1}^{N} a_i^*(h(t-1))\right] - \left[a_i^1(h(t-2)) \sum_{i=1}^{N} a_i^*(h(t-1))\right] \quad (5)$$

Using $h(t) = \sum_{j=1}^{m} b(t-j+1) 2^{j-1}$ and inserting (5) in (4), we obtain

$$\left.\frac{db}{dt}\right|_{t+1} = \delta\left(\sum_{i=1}^{N} a_i^*(h(t))\right) \sum_{i=1}^{N} \left\{ \delta\left(q_i(h(t))\right) \sum_{i=1}^{N} a_i^*(h(t-1)) \right.$$

$$\left[a_i^2\left(\sum_{j=1}^{m} b(t-j-1) 2^{j-1}\right) - a_i^1\left(\sum_{j=1}^{m} b(t-j-1) 2^{j-1}\right)\right] \times$$

$$\left[a_i^2\left(\sum_{j=1}^{m} b(t-j+1) 2^{j-1}\right) - a_i^1\left(\sum_{j=1}^{m} b(t-j+1) 2^{j-1}\right)\right] +$$

$$\left. \Theta\left(q_i(h(t))\right) \frac{\delta a_i^2\left(\sum_{j=1}^{m} b(t-j+1) 2^{j-1}\right)}{\delta t} + \left[1 - \Theta\left(q_i(h(t))\right)\right] \frac{\delta a_i^1\left(\sum_{j=1}^{m} b(t-j+1) 2^{j-1}\right)}{\delta t} \right\} \quad (6)$$

If $\sum_{i=1}^{N} a_i^*(h(t-1))$, $\sum_{i=1}^{N} a_i^*(h(t-2))$, ..., $\sum_{i=1}^{N} a_i^*(h(t-m))$ all have the same sign, the right-hand-side of (6) becomes 0, thus proving that a constant bit $b(t)$, corresponding to either an exponential increase or decrease in price, is a Nash equilibrium.

## 3 Introducing decoupled strategies to determine the onset of convergence towards Nash-equilibria in multi-period market games: theoretical framework

As mentioned, the mere size of the strategy space for one-shot games is $2^{2^M}$, which considerably complicates a proper understanding of simple market models like the minority game and the $-game, despite the simplicity of their payoff functions. It is therefore out of the question to explore the full strategy space in order to gain insights into the way people would react, even in simple market games like this. We propose instead to concentrate on a certain subclass of strategies.

Let us call $s_i(t \mid \vec{h}_M(t))$ the action of strategy $s_i$ at time t, *conditioned* on observing a given price history, $\vec{h}_M(t)$, at time t over the last M time steps. $\vec{h}_M(t)$ is a binary string of -1's and +1's describing the last M directions of price movements observed at time t. We now note that some strategies will be *independent* of $\vec{h}_M(t)$ over the next L time steps. That is, whatever the price history, over the next t + Q time steps, the strategy $s_i(t + Q)$ will always issue the same prediction *independently* of the price history between t and t+Q. We call such strategies Q-time-steps *decoupled* (*Andersen and Sornette, 2005*). The simplest example of a strategy that is decoupled is the strategy that always issues a buy (sell) action, independently of the past price history $\vec{h}_M(t)$. Such a strategy is trivially an infinite-number-of-time-steps decoupled. However, the probability that any player will hold this specific strategy is very small indeed, with a probability that goes as $\frac{S}{2^{2^M}}$.

As will be shown in the following, it is advantageous to split the order balance in two, so that it can be written in terms of decoupled and coupled strategies:

$$A(t)^{\vec{h}_M} = A(t)^{\vec{h}_M}_{coupled} + A(t)^{\vec{h}_M}_{decoupled}$$

We have included the superscript $\vec{h}_M$ to emphasize that the order imbalance is conditioned on observing the price history $\vec{h}_M$ at time t.

The size of $A(t)^{\vec{h}_M}_{decoupled}/N$ then gives the percentage of decoupled strategies at time t, and will be used as a predictor for the actions of the participants. In the extreme case where the condition

$$|A(t)^{\vec{h}_M}_{decoupled}| > N/2$$

is fulfilled, the prediction becomes *certain*, since the action at time t+2 of more than half the population takes the same sign (buy/sell) *independently* of what happens at time t+1. The greater the value of $A(t)^{\vec{h}_M}_{decoupled}/N$, the better we should be able to predict the action at time t+2. This fact will be used in experiments where we try to predict the actions of the participants. For a discussion of the effect of group size N, see also *(Nosenzo et al., D. 2015)*.

**4 Laboratory experiments**

We performed a series of 10 experiments at the Laboratory of Experimental Economics in Paris (LEEP). The experiments ran over 60 periods. In each period, the students received general economic news and could decide whether to buy or sell an asset, or simply do nothing. At the end of the 60 periods, the students were paid pro rata according to their performance (for more details about the way the experiments were set up, see Appendix B).

At the beginning of the experiment, the students were told that the asset was, at this initial stage, properly priced according to rational expectations (*Fama 1970; Muth 1961*). This meant that only information regarding changes in the dividends on the asset or interest rates should have a direct influence on the price of the asset (for an interesting study with varying fundamental values, see, e.g., *Stockl et al. 2015)*. The information flow consisted of general news from real past records of Bloomberg news items. News was selected in such a way that the general trend over the 60 consecutive periods was neutral. Then, according to rational expectations, there should be no overall price movement of the asset at the end of the 60 time periods. The price was thus expected to oscillate around the fundamental value throughout the experiment.

**5 Results**

Figures 1-5 show the price history versus time for the ten experiments (E1-E10), as well as the decoupling parameter versus time obtained via Monte Carlo (MC) simulations. Each figure represents the data and simulation from two experiments. The MC simulations "slaved" the price history from the experiments as input to the agents in the $G simulations. That is, instead

of having the agents reacting to their own repeated-game actions, the agents in the simulations would instead use the actions (price history) of the participants. Each $G MC simulation was done with a fixed number of agents N=10 (number of participants in the experiments), but randomly generated initial strategies. The number of strategies, S, used by the agents and the memory, M, were also randomly generated S $\epsilon$ [1,$S_{max}$], M $\epsilon$ [1, $M_{max}$] , with $M_{max}$=6 and $S_{max}$=10 reflecting the maximum values of memory and number of strategies thought to be used by the participants in the experiments (we checked this assumption by interviewing the participants after the experiments). Simulations with larger values of $S_{max}$ and $M_{max}$ were performed, showing similar trends, as presented in the following. In total L=1000 MC simulations were performed for each experiment.

The topmost plot of Fig. 1 shows the price evolution as a function of time for experiment E1. The second plot from the top shows the percentage of decoupled strategies used by the agents ($d^+$ for decoupled strategies recommending buy, represented by a solid line, and $d^-$ for decoupled strategies recommending sell, represented by a dashed line) as a function of time. The percentage for each time period t is obtained by averaging over all MC samples.

Insert figure 1 around here

**Fig. 1.** Experiments E1 and E2. First and third plot: price evolution in experiments E1 and E2, respectively. Second (E1) and fourth (E2) plots: percentage of decoupled strategies used by the agents. Solid line: $d^+$ for decoupled strategies recommending buy as a function of time. Dashed line: $d^-$ for decoupled strategies recommending sell as a function of time.

We now use the difference $\Delta D(t) \equiv |d^+(t)-d^-(t)|$ as a predictor of the action of the participants in the next time step t+1. Table 1 shows $\Delta D$ for values greater than 0.2, the success rate in predicting the next price direction, and the number of periods used in the calculation of the success rate. The experiment E1 corresponds to the simplest case of all the 10 experiments, since in this case the price dynamics created by the participants shows the formation of a clear financial bubble. This is captured by the split in decoupled strategies measured by $\Delta D$ leading to a 100% success rate. However, this is less trivial than it appears since, as noted by *Roszczynska et al. (2012)*, decoupling of strategies is *sufficient* but *not necessary* for financial bubble formation. In other words, the onset of speculation is possible without a split developing between $d^+(t)$ and $d^-(t)$. For a more complete discussion of this point, see *Roszczynska et al. (2012)*.

Insert table 1 around here

The lower two plots of Fig. 1 show the price evolution and corresponding percentage of decoupled strategies for experiment E2. Once again, we see a tendency for speculative action to develop, creating a financial bubble in prices, but a factor of 10 less than in E1. Using *ΔD* as predictor, we are again able to predict the actions of the participants with a high success rate (see Table 2).

Insert table 2 around here

Figure 2 shows the price evolution and corresponding percentage of decoupled strategies for experiments E3 and E4. In E3, a very weak overall trend develops, whereas the first part of E4 is trendless, followed by a weak upward trend, and ending with a small downward trend. Using *Δ*D as predictor, we are again able to predict the actions of the participants with a rather high success rate (see Tables 3 and 4). It should be noted that the first half of E4 is trendless, but *Δ*D is still large, as can be seen from Figure 2, and, a reliable predictor even in this trendless case, as can be seen from Table 4.

Insert figure 2 around here

**Fig. 2.** Experiments E3 and E4. First and third plot: price evolution of experiments E3 and E4, respectively. Second (E3) and fourth (E4) plots: percentage of decoupled strategies used by the agents. Solid line: $d^+$ for decoupled strategies recommending buy as a function of time. Dashed line: $d^-$ for decoupled strategies recommending sell as a function of time.

Insert table 3 around here

Insert table 4 around here

Insert table 5 around here

Figure 3 shows the price evolution and corresponding percentage of decoupled strategies for experiments E5 and E6. Both experiments are more or less without trends, with the price fluctuating around the fundamental value. A small split in $d^+(t)$ and $d^-(t)$ is only seen for initial times in E5, whereas a split exists throughout E6. For E5, $\varDelta D$ never exceeds 0.2 and no prediction can be made. For E6, $\varDelta D$ only exceeds the threshold value 0.2 in a few time periods, but the few predictions made appear to be slightly better than random or random (see Table 6).

Insert figure 3 around here

**Fig. 3.** Experiments E5 and E6. First and third plot: price evolution of experiments E5 and E6, respectively. Second (E5) and fourth (E6) plots: percentage of decoupled strategies used by the agents. Solid line: $d^+$ for decoupled strategies recommending buy as a function of time. Dashed line: $d^-$ for decoupled strategies recommending sell as a function of time.

Insert table 6 around here

Figure 4 shows the price evolution and corresponding percentage of decoupled strategies for experiments E7 and E8. Weak trends develop in the price formation, but not in the split of $d^+(t)$ and $d^-(t)$. For these two experiments, the days on which we can issue a prediction are too few to have any meaningful statistical significance.

Insert figure 4 around here

**Fig. 4.** Experiments E7 and E8. First and third plot: price evolution of experiments E7 and E8 respectively. Second (E7) and fourth (E8) plots: percentage of decoupled strategies used by the agents. Solid line: d$^+$ for decoupled strategies recommending buy as a function of time. Dashed line: d$^-$ for decoupled strategies recommending sell as a function of time.

Insert table 7 around here

Insert table 8 around here

Figure 5 shows the price evolution and corresponding percentage of decoupled strategies for experiments E9 and E10. There is no trend present in E9, whereas a very weak trend develops in E10. A small number of splits $\Delta D$ develop in E9, more in E10, but in neither case can predictions be made.

Insert figure 5 around here

**Fig. 5.** Experiments E9 and E10. First and third plot: price evolution of experiments E9 and E10, respectively. Second (E9) and fourth (E10) plots: percentage of decoupled strategies used by the agents. Solid line: d$^+$ for decoupled strategies recommending buy as a function of time. Dashed line: d$^-$ for decoupled strategies recommending sell as a function of time.

Insert table 9 around here

Insert table 10 around here

In order to test the stability of the Monte Carlo simulations, we performed a series of bootstrap simulations. Figure 6 shows the bootstrap simulations performed on the data for the experiments E1, E2, and E3. Similar bootstrap simulations were performed for E4 and E5, with the same conclusions as noted in the following for E1, E2, and E3. The figures were obtained by doing 100 Monte Carlo simulations for each experiment and calculating $d^+$ and $d^-$. For each of the 100 Monte Carlo simulations, 100 more replica Monte Carlo simulations were then performed in order to be able to assign 90% and 10% confidence levels. The results are presented in Fig. 6, where the solid lines represent a MC realisation of $d^+$ and $d^-$, while the dashed lines give the 90% and 10% confidence levels calculated using the 100 replica MC simulations, i.e., in only 10% of the cases would the replica MC simulations find a value of $d^+$ (or $d^-$) below what was found by the MC simulation, and in 90% of the cases the replica MC simulations would find a value of $d^+$ (or $d^-$) not exceeding what was found by the MC simulation. As can be seen, the levels of $d^+$ and $d^-$ in the MC simulations are very stable, confirming the stability of our predictions obtained via the split in $d^+$ and $d^-$.

To summarize: We have made a series of 10 experiments, each ran over 60 time periods, in which students received general economic news, and could at each time period decide whether to buy or sell an asset. Some of the experiments showed the tendency to price bubble formation, whereas others fluctuated around the (constant) fundamental value of the asset. After the performance of the experiments, the price time series generated were used as input to agent based MC simulations. By studying the difference of decoupled strategies of the MC agents, $\Delta D(t) \equiv |d^+(t) - d^-(t)|$, we were able to issue a prediction of the action of the participants in the next time step t+1. In 5 out of the 10 experiments we were not able to use our method to issue predictions. However in 5 out of the 10 experiments we were able over several time periods to issue predictions out-of-sample of the behavior of the participants. In 5 of these 4 experiments, we obtained very high success rates, most of the events predicted with 65% succes or more. In one experiment our predictions were mostly random. Finally we used a bootstrap simulation to verify that our predictions were stable, and independent of the parameters used in the MC simulations. Overall our technique of using decoupled strategies appears to be a promising way to probe the heuristics in participants in financial market games.

Insert figure 6 around here

**Fig. 6.** Bootstrap simulations performed on the data for the experiments E1, E2, and E3. The figures were obtained by doing 100 Monte Carlo simulations for each experiment and calculating $d^+$ and $d^-$. For each of the 100 Monte Carlo simulations, 100 more replica Monte Carlo simulations were then performed in order to be able to assign 90% and 10% confidence levels. Solid lines represent the result for a MC realization of $d^+$ and $d^-$, while dashed lines give the 90% and 10% confidence levels calculated using the 100 replica MC simulations, i.e., in only 10% of the cases would the replica MC simulations find a value of $d^+$ (or $d^-$) below what was found by the MC simulation, and in 90% of the cases the replica MC simulations would find a value of $d^+$ (or $d^-$) not exceeding what was found by the MC simulation.

**6 Conclusion**

We have introduced a new methodology to describe the onset of convergence towards Nash equilibria in simple repeated-games with infinitely large strategy spaces. This was shown to enable a better understanding of the heuristics used in decision-making. The method works by restricting to a special finite subset of strategies which we called decoupled strategies. We performed 10 multi-period simple financial market experiments, and applied our methodology within the framework of a simple financial market game called the $-game.

We have illustrated how the method was able to predict (at certain special times) participants' actions with a high success rate. We note that the decoupled strategy methodology is not limited to the $-game presented in this article, but can be applied in the context of any multi-period game.

## Appendix A: Implementation of the experiments

The number of participants in each experiment was fixed at 10. There was only one asset in our financial market that participants could either buy or sell, and short selling was allowed. The initial price of the asset was fixed at 5 euros with an expectation of a 10 cent dividend payout at the end of 60 time periods. Each of the 60 time periods lasted 15 seconds. In each time period, participants were presented with brief statements of economic news and they could either buy or sell ONE asset or do nothing. The participants were told that the asset was correctly priced according to rational expectations (*Muth, 1961*), that is, the price of the asset was supposed to correctly reflect all future discounted cash flow accrued to the asset. The participants could, at zero interest rate, borrow money to buy shares, and short selling was allowed. The general financial information was taken from real financial news items obtained on Bloomberg over a two-week time period. Students were told the asset represented a portfolio of assets like an ETF or an index. They were all simultaneously presented with the same information, meant to reflect general financial news, e.g., good or bad US employment figures, commodity price changes, etc. The news items were the same in all experiments and were chosen without positive or negative bias.

At the end of each time period, the participants' orders were gathered and a new market price calculated, based on the order imbalance (with sign and magnitude determining the direction and size of the price movement) (*Holthausen et al. 1987*). That is, the price at time t was given by $P(t) = P(t-1)e^{A(t)/b}$, where A(t) is the order imbalance at time t and b the liquidity of the market (in the experiments chosen as b≡ 10 ∗ N, representing a market crash/bubble of 10 percent when all N participants chose the same action). This was then shown to the participants graphically on their computer screen. Throughout the experiment, participants had a continuous update of the number of shares held and their gains/losses.

At the end of each experiment, a pool of 200 euro was distributed pro rata among the participants who had made a gain.

FIGURE 1

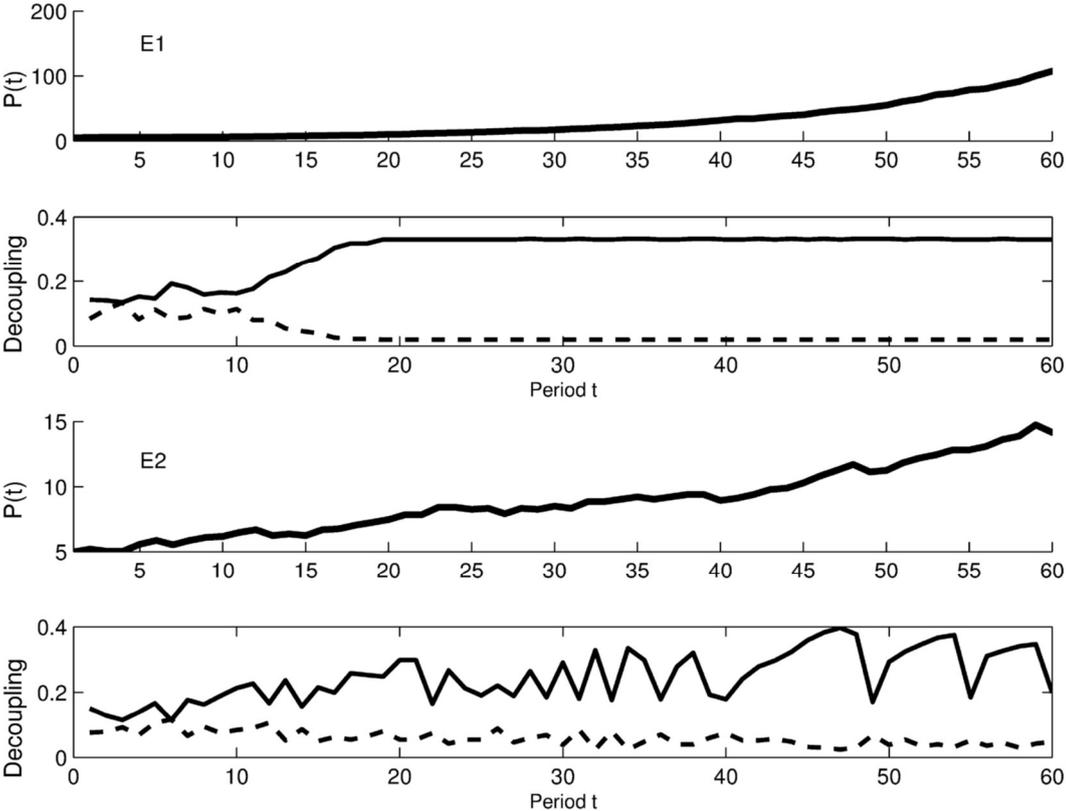

FIGURE 2

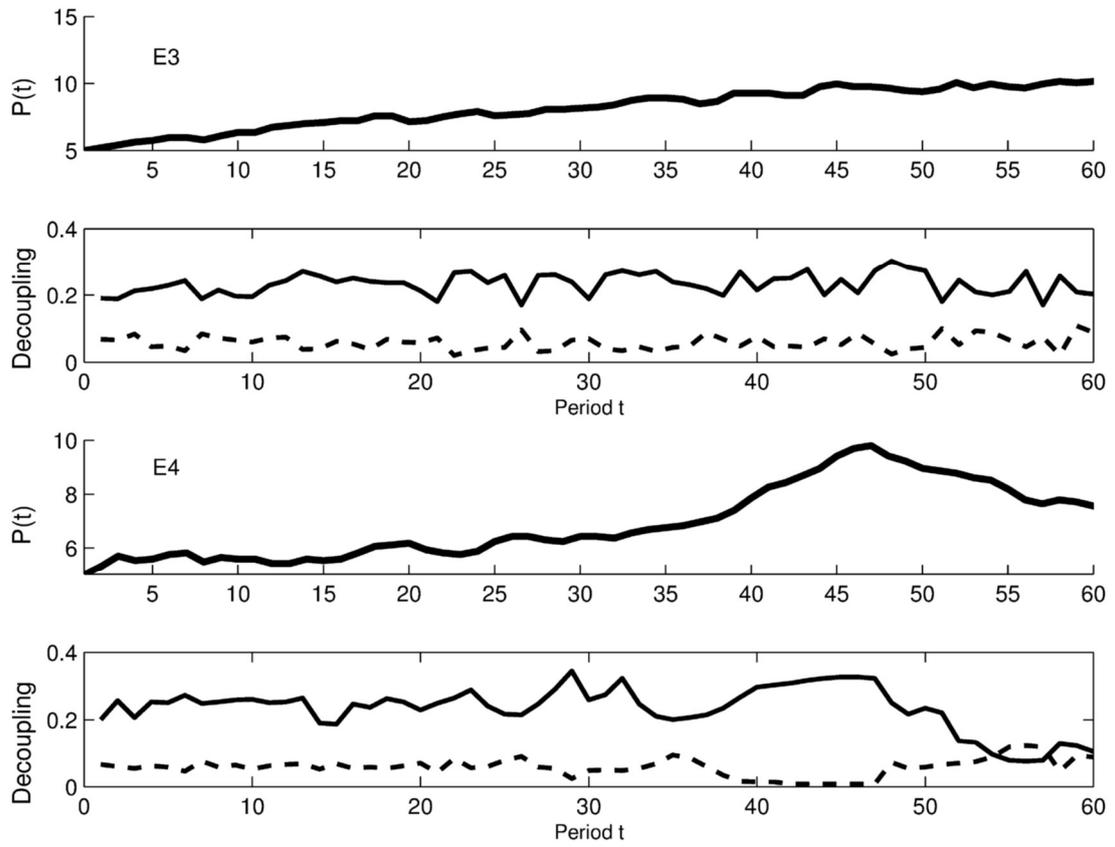

FIGURE 3

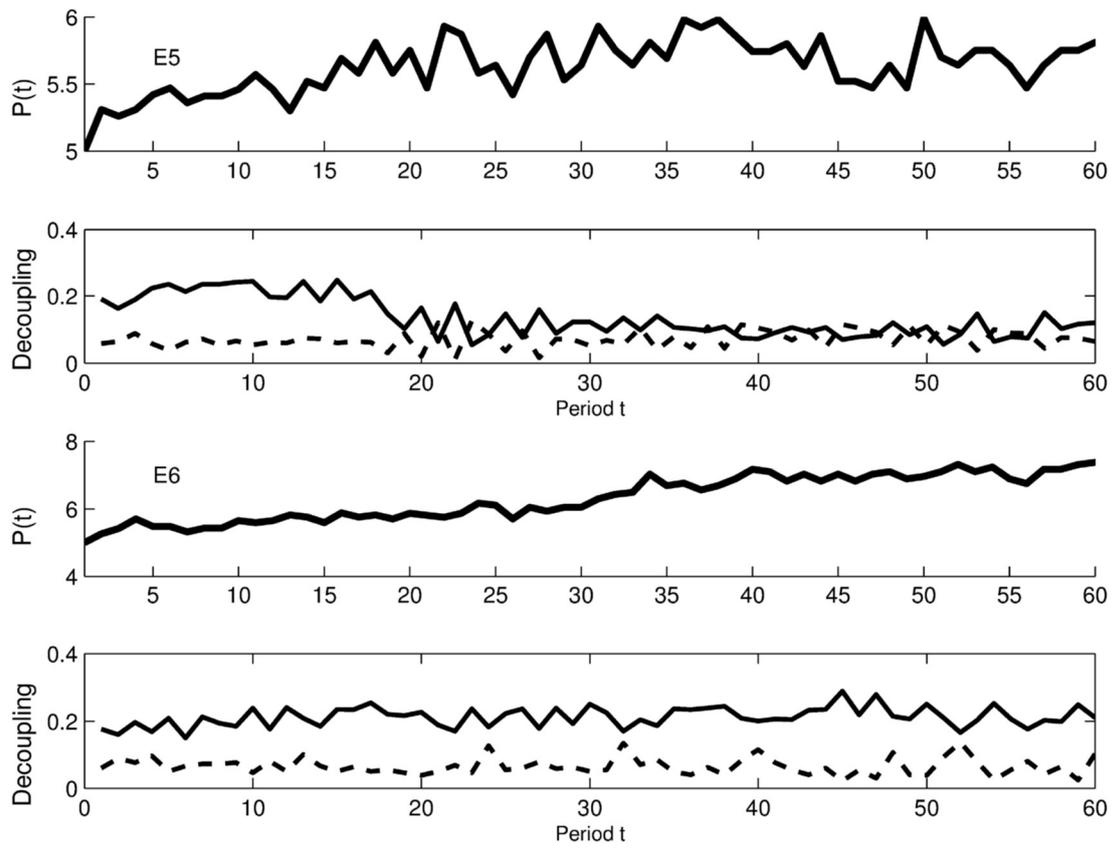

FIGURE 4

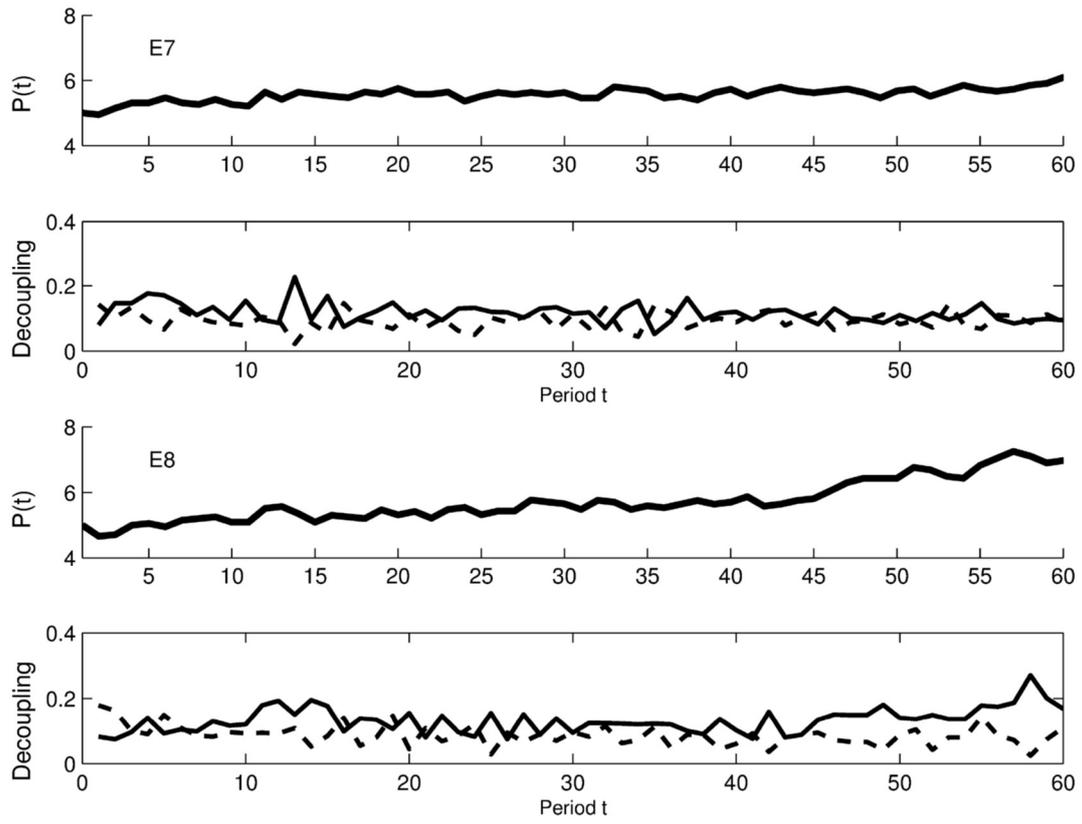

FIGURE 5

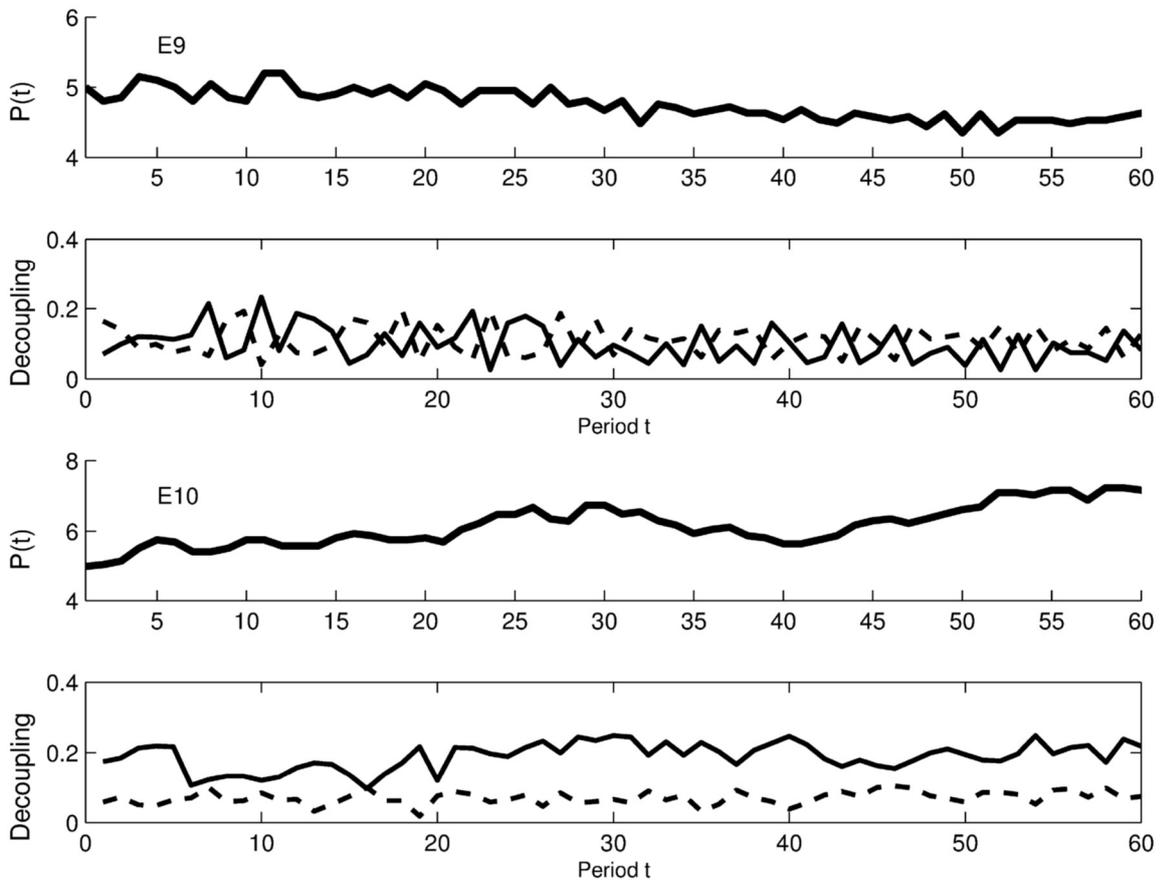

FIGURE 6

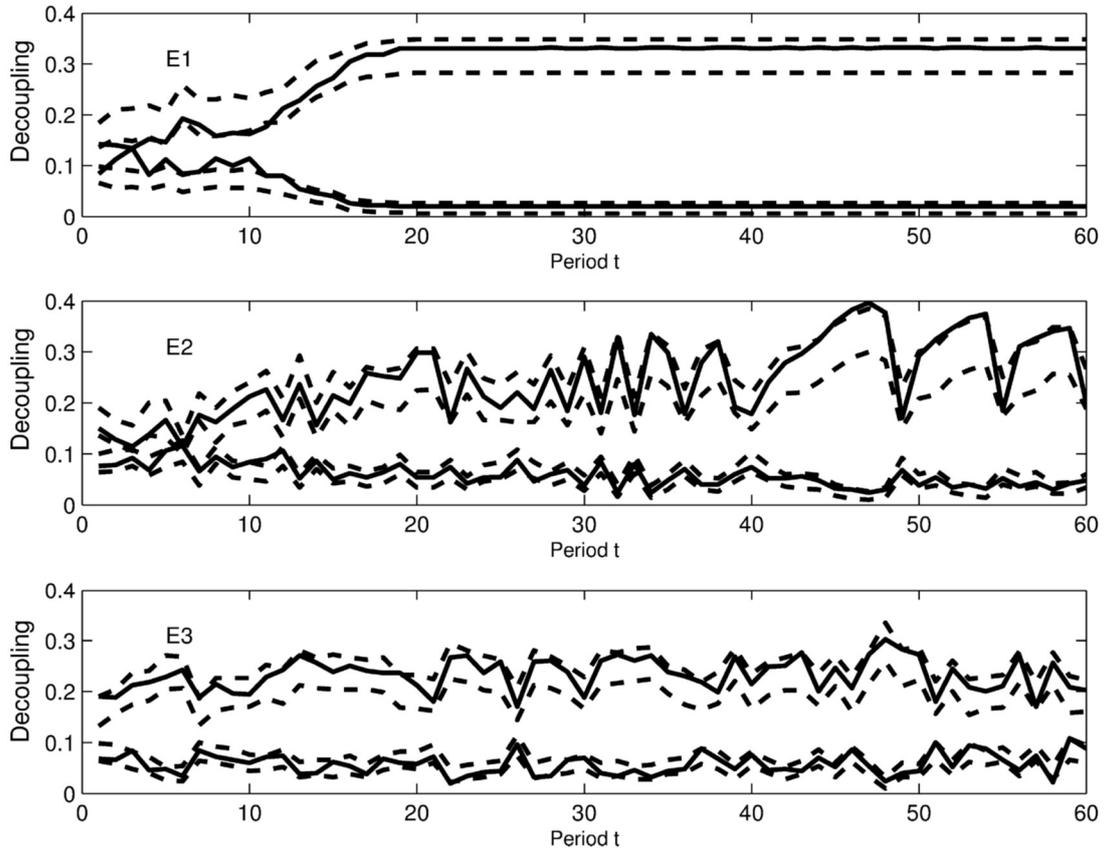

**Table 1**

ΔD , success rate, and corresponding number of events used to calculate the success rate for experiment E1.

| ΔD | Success rate | Number of events |
|---|---|---|
| 0.2 | 1.0 | 46 |
| 0.22 | 1.0 | 45 |
| 0.24 | 1.0 | 44 |
| 0.26 | 1.0 | 44 |
| 0.28 | 1.0 | 43 |
| 0.3 | 1.0 | 41 |
| 0.32 | - | 0 |

**Table 2**

ΔD , success rate, and corresponding number of events used to calculate the success rate for experiment E2.

| ΔD | Success rate | Number of events |
|---|---|---|
| 0.2 | 0.70 | 27 |
| 0.22 | 0.68 | 25 |
| 0.24 | 0.73 | 22 |
| 0.26 | 0.69 | 16 |
| 0.28 | 0.62 | 13 |
| 0.3 | 0.64 | 11 |
| 0.32 | 0.67 | 6 |
| 0.34 | 0.75 | 4 |
| 0.36 | - | 0 |

**Table 3**

ΔD , success rate, and corresponding number of events used to calculate the success rate for experiment E3

| ΔD | Success rate | Number of events |
|---|---|---|
| 0.2 | 0.57 | 23 |
| 0.22 | 0.67 | 15 |
| 0.24 | 0.67 | 3 |
| 0.26 | - | 0 |

## Table 4

ΔD , success rate, and corresponding number of events used to calculate the success rate for experiment E4

| ΔD | Success rate | Number of events |
|---|---|---|
| 0.2 | 0.65 | 20 |
| 0.22 | 0.73 | 15 |
| 0.24 | 0.73 | 11 |
| 0.26 | 0.70 | 10 |
| 0.28 | 0.67 | 9 |
| 0.30 | 0.57 | 7 |
| 0.32 | - | 0 |

## Table 5

ΔD , success rate, and corresponding number of events used to calculate the success rate for experiment E5

| ΔD | Success rate | Number of events |
|---|---|---|
| 0.2 | - | 0 |

## Table 6

ΔD , success rate, and corresponding number of events used to calculate the success rate for experiment E6

| ΔD | Success rate | Number of events |
|---|---|---|
| 0.2 | 0.63 | 8 |
| 0.22 | 0.50 | 4 |
| 0.24 | 0.50 | 2 |
| 0.26 | - | 0 |

**Table 7**

ΔD , success rate, and corresponding number of events used to calculate the success rate for experiment E7

| ΔD | Success rate | Number of events |
|---|---|---|
| 0.2 | 0.0 | 1 |
| 0.22 | - | 0 |

**Table 8**

ΔD , success rate, and corresponding number of events used to calculate the success rate for experiment E8

| ΔD | Success rate | Number of events |
|---|---|---|
| 0.2 | 1.00 | 1 |
| 0.22 | - | 0 |

**Table 9**

ΔD , success rate, and corresponding number of events used to calculate the success rate for experiment E9

| ΔD | Success rate | Number of events |
|---|---|---|
| 0.2 | - | 0 |

**Table 10**

ΔD , success rate, and corresponding number of events used to calculate the success rate for experiment E10

| ΔD | Success rate | Number of events |
|---|---|---|
| 0.2 | - | 0 |